# Inadmissible Class of Boolean Functions under Stuck-at Faults

*Debesh K. Das[1], Debabani Chowdhury[1], Bhargab B. Bhattacharya[2], Tsutomu Sasao[3]*
[1]Computer Sc. & Engg. Dept., Jadavpur University, Kolkata -700 032, India
[2]ACM Unit, Indian Statistical Institute, Kolkata 700 108, India
[3]Dept. of Computer Science, Meiji University, Kawasaki, Kanagawa 214-8571, Japan
{debeshkdas@gmail.com, bhargab.bhatta@gmail.com, sasao@cs.meiji.ac.jp}

***Abstract:*** Many underlying structural and functional factors that determine the fault behavior of a combinational network, are not yet fully understood. In this paper, we show that there exists a large class of Boolean functions, called *root functions*, which can never appear as faulty response in irredundant two-level circuits even when any arbitrary multiple stuck-at faults are injected. Conversely, we show that any other Boolean function can appear as a faulty response from an irredundant realization of some root function under certain stuck-at faults. We characterize this new class of functions and show that for *n* variables, their number is exactly equal to the number of independent dominating sets (Harary and Livingston, *Appl. Math. Lett.,* 1993) in a Boolean *n*-cube. We report some bounds and enumerate the total number of root functions up to 6 variables. Finally, we point out several open problems and possible applications of root functions in logic design and testing.

***Keywords:*** Boolean functions, hypercube, redundancy, stuck-at faults, testing.

## 1. Introduction

Although test generation and design-for-testability (DFT) issues have been studied extensively over a few decades, characterization of the impossible class of faulty functions (ICFF) [1], i.e., the set of Boolean functions which can never appear as faulty response in a combinational circuit, is not yet well understood. Several structural factors, e.g., choice of gates, network interconnection and fanout geometry, number of logic levels, factorization, multiplicity of outputs, redundancy, modularity, technology used, and fault model, determine the output functions in the presence of faults. Some early results regarding fault behavior under stuck-at faults were reported by Hayes [2], [3], Fujiwara [4], and by Das et al. [5]. It is known that no single stuck-at fault in a multi-output irredundant network realizing a non-constant function can change it to its complement, and the result is conjectured to be true in any single-output irredundant circuits even for multiple faults [3]. Some characterization of ICFF under short-circuit faults appeared earlier in the literature [1]. Also, realizations of different classes of Boolean functions such as fanout-free and unate functions [6,12,13], elementary functions [7], threshold functions [8, 12], symmetric functions [9, 12], bent functions [10], linear functions and index generation functions [11] may exhibit special properties in faulty responses. For instance, the set of functions observed in unate and fan-out-free circuit realizations in the presence of stuck-at faults, is known to satisfy certain closure properties [4]. A few other interesting functional properties under stuck-at faults in combinational and sequential circuits were reported earlier [14,15,16]. However, because of multitude of circuit realizations, even partial characterization of ICFF seems to be a very complex problem.

In this paper, we introduce a new class of Boolean functions, called as *root functions*, and show that no such function can appear as a faulty response in any two-level irredundant AND-OR circuits under stuck-at faults. We prove that set of all root functions and the set of their possible faulty response functions are not only mutually exclusive but their union also collectively exhausts all $2^{2^n}$ Boolean functions of *n* variables for any *n*. More interestingly, we show that the number of such root functions is exactly equal to the number of independent domination in a hypercube of dimension *n* [17].

## 2. Preliminaries

Let $F(x_1, x_2, x_3,...., x_n)$ be a switching function of *n* variables. A *literal* is a variable or its complement. A *minterm* is a product of literals in which every literal appears once [12]. A minterm (*m*) is *true (false) minterm* if $F(m) = 1 (0)$. The *weight of a minterm* is the number of uncomplemented variables that appear in it. The *Hamming distance between two minterms* is the number of bits in which they differ. A minterm is said to be adjacent to another minterm, if their Hamming distance is unity. A Boolean function may be defined by the set of true minterms and can be expressed as a sum-of-products (s-o-p) form. An *n*-variable Boolean function *F* is said to be *non-vacuous* if *F* cannot be expressed with fewer than *n* variables.

*Definition 1*: [12] A sum-of-products (s-o-p) expression of a Boolean function is called *irredundant*, if no term or literal can be deleted from the expression without altering the function.

An *implicant* is a product of literals covering of one or more true minterms. A *prime implicant* of a function is an implicant that cannot be covered by another implicant with a fewer literals. A Boolean function can be pictorially represented using a *Karnaugh map (K-map)* for small values of $n$ [12].

*Definition 2*: A minterm $m_i$ is said to *dominate* another minterm $m_j$, if $m_j$ is adjacent to $m_i$. A false minterm $m_i$ is said to be *essentially dominated* if there exists only one true minterm which dominates $m_i$.

*Example 1*: Consider the *K*-map shown in Figure 1. The true minterm ($x_3x_2x_1$) 000 is dominating three false minterms (001, 010, 100), whereas the true minterm 101 is dominating three false minterms (001, 111, 100). The false minterm 011 is not dominated by any other true minterm. The minterms 010 and 111 are essentially dominated.

Figure 1: Minterm dominance

*Remark 1:* In any $n$-variable Boolean function, a minterm always dominates $n$ minterms, and conversely, every minterm is dominated by $n$ minterms.

*Definition 3:* [13] Two functions are said to be *equivalent* if one can be transformed to the other by (i) negation (*N*) of one or more input variables, and/or (ii) permutation (*P*) of two or more input variables. The functions that are equivalent under operation (i) form *N-equivalent* class and those under operation (ii) form *P-equivalent* class, and those equivalent under operations (i) and/or (ii) form *NP-equivalent* class.

## 2. Root Functions

In this section, we present new properties of root functions in the context of fault behavior, results on their bounds, and their relationship to hypercubes.

### 2.1 Preliminaries and properties

*Definition 4* [15]: A Boolean function *F* is said to be *isolated* if every true minterm of *F* is a prime implicant by itself, *i.e.*, no two true minterms of *F* are adjacent.

A function *F* is called maximal if for every false minterm $m_i$ of *F*, there exists at least one true minterm which dominates $m_i$.

*Definition 5* [5]: A Boolean function *F* is a *root function* if *F* is non-vacuous, isolated and maximal.

*Example 2*: In Boolean functions of two variables, there are only two root functions, shown in Figure 2.

Figure 2: Two root functions for $n = 2$

Figure 3: Two root functions for $n = 3$

*Example 3:* Fig. 3 shows two examples of root functions of three variables. However, there exist four other root functions for $n = 3$, which will be discussed later.

*Lemma 1:* Let $N(R, n)$ be the number of true minterms in a root function $R(n)$ with $n$-variables. Then $N(R, n) \leq 2^{n-1}$.

*Proof:* Consider the *K*-map for an *n*-variable Boolean function. There are $2^n$ cells in the map. We partition the cells into $2^{n-1}$ pairs consisting of two adjacent cells each. Let us construct a function *F* such that its true minterms correspond to every alternate cell of the map. Thus, there will be $2^{n-1}$ true minterms in *F*, each placed in one pair. Hence *F* is isolated and maximal, *i.e.*, a root function $R(n)$ of *n* variables. If we add an extra true minterm in *F*, by pigeon-hole principle [20], there will be at least one pair of adjacent cells containing two true minterms. Hence, *F* will no longer remain isolated and cannot be a root function. Therefore, $N(R, n) \leq 2^{n-1}$. □

*Lemma 2:* $N(R, n) \geq \lceil 2^n/(n+1) \rceil$.

*Proof:* As observed in Remark 1, $N(R, n)$ true minterms can dominate at most $n \times N(R, n)$ false minterms, *i.e.*, $N(R, n) + n.N(R, n) \geq 2^n$. Hence, $N(R, n) \geq \lceil 2^n/(n+1) \rceil$. □

*Definition 6*: A root function that contains the maximum number of true minterms is called *max-root*. A root function that contains the minimum number of true minterms is called *min-root*.

*Corollary 1*: For any $n$, there exist exactly two max-root functions $R_1$ and $R_2$ such that $N(R_1, n) = N(R_2, n) = 2^{n-1}$.

This follows from the construction used in the proof of Lemma 1. Clearly, $R_1 = x_1 \oplus x_2 \oplus x_3 \oplus \ldots \oplus x_{n-1} \oplus x_n$ and $R_2 = 1 \oplus x_1 \oplus x_2 \oplus x_3 \oplus \ldots \oplus x_{n-1} \oplus x_n$ are two max-root functions. In other words, the parity function and its complement are the only max-root functions. Also, the upper bound as stated in Lemma 1 is tight.

*Example 4*: From Lemma 2, it follows that $N(R, n) \geq 2$, 2, 4, 6, 10 and 16 for $n =$ 2, 3, 4, 5, 6 and 7 respectively. However, the lower bound is not always achievable for min-root functions. For $n = 2$, each of the functions shown in Figure 2(a) and 2(b) is a min-root and a max-root. For $n = 3$, it is easy to check that the function shown in Figure 3(a) is a min-root, whereas that in Figure 3(b) is a max-root. For $n = 4$, Figure 4 shows a min-root function where the lower bound on $N(R, n)$ is satisfied as $N(R, 4) \geq 4$. However, we will show next that for $n = 5$, a min-root function will consist of at least 8 true minterms, and thus the lower bound of $N(R, 5) \geq 7$ is not achievable.

*Lemma 3*: If $R_1(n)$ be a root function $n$ variables such that $N(R_1, n) = k$. Then any function $R_2(n)$ which is obtained by complementing a literal in $R_1(n)$ is also a root function. In other words, if $R_2(n)$ is $N$-equivalent to $R_1(n)$, then $R_2(n)$ is a root function implying $N(R_1, n) = N(R_2, n) = k$.

*Proof*: The function $R_1(n)$ can be expressed in the form $R_1(n) = f_1 x_i + f_2 x'_i$ with respect to a literal $x_i$ of the function, where $f_1$ and $f_2$ are independent of $x_i$. Since a root function in non-vacuous by definition, $f_1 \neq f_2$. Let us consider an $N$-equivalent function of $R_1(n)$ which is obtained by complementing $x_i$: $R_2(n) = f_2 x_i + f_1 x'_i$. It is now easy to show that $R_2(n)$ is different from $R_1(n)$ and that it is a root function. □

*Theorem 1*: Let $R(n)$ be a root function of $n$ variables. If $N(R, n) = k$, then there exists a root function $R(n + 1)$ of $(n+1)$ variables such that $N(R, n+1) = 2k$.

*Proof:* Consider a root function $R(n)$ of $n$ variables ($x_1, x_2, x_3, \ldots x_i, \ldots, x_n$). By Lemma 1, $k \leq 2^{n-1}$. Let $R_i(n)$ be a function which is obtained by complementing $x_i$ in $R(n)$. By Lemma 3, $R_i(n)$ is also a root function. Let us now construct an $(n+1)$-variable function $R(n + 1)$ $x'_{n+1}.R(n) + x_{n+1}.R_i(n)$. As $R_i(n)$ is obtained by complementing $x_i$ in $R(n)$, the Hamming distance between a true minterm in $x'_{n+1}.R(n)$ and that in $x_{n+1}.R_i(n)$ must be at least 2. Since, both $R(n)$ and $R_i(n)$ are root functions, $R(n + 1)$ is a root function of $(n+1)$ variables and $N(R, n+1) = 2k$. □

*Example 5*: Consider the root function in Figure 3(a) with two true minterms for $n = 3$. Based on this, we can construct a root function with four true minterms (0000, 0111, 1101, 1010) for $n = 4$, as shown in Figure 4. Similarly, we can construct a root function of five variables with eight true minterms (00000, 00111, 01101, 01010, 10001, 10110, 11100, 11011).

| $x_4x_3$ \ $x_2x_1$ | 00 | 01 | 11 | 10 |
|---|---|---|---|---|
| 00 | 1 |   |   |   |
| 01 |   |   | 1 |   |
| 11 |   | 1 |   |   |
| 10 |   |   |   | 1 |

Figure 4: A root function for $n = 4$

*Remark 2*: Although by Lemma 2, we have $N(R, 5) \geq 6$, for $n = 5$, the lower bound is not achievable. It is easy to prove from essential domination argument that the minimum number of true minterms for a root function of 5 variables is 8, an example of which is given earlier.

*Example 6*: A root function for $n = 6$ variables can be constructed by using Theorem 1 on the above example for $n = 5$. This yields a root function with 16 true minterms: (000000, 000111, 001010, 001101, 010001, 010110, 011011, 011100, 100001, 100110, 101011, 101100, 110000, 110111, 111010, 111101). However, a min-root function for $n = 6$ consists of 12 true minterms: (000000, 000111, 001100, 011010, 011001, 010100, 101011, 100110, 100101, 110011, 111000, 111111). Thus, for $n = 6$, the lower bound for min-root function is not achievable.

*Example 7*: For $n = 7$, the lower bound is achievable as there exists a min-root of 16 true minterms: (0001 011, 0001 100, 0010 010, 0010 101, 0111 000, 0111 111, 0100 001, 0100 110, 1000 000, 1000 111, 1011 001, 1011 110, 1110 011, 1110 100, 1101 010, 1101 101).

**Root functions other than min- or max-root**

Lemma 1 and Lemma 2 provide an upper and a lower bound on the number $N(R, n)$ of true minterms in a root function $R(n)$ for $n$-variables. While the upper bound is always achievable for any $n$, the lower bound may not always be achievable as demonstrated earlier by several examples. Here, we show that even between two achievable bounds not every value is feasible for a root function. For example, $N(R, 4) = 4$ for a min-root (Figure 4) and $N(R, 4) = 8$ for a max-root (4-variable parity function). However, for $n = 4$, no root function exists with 6 or 7 true minterms. We prove this in the following lemma.

*Lemma 4*: For $n = 4$, no root function with 6 or 7 true minterms exists.

*Proof*: Let us draw the *K*-map for four variables in two rows and 8 columns as shown in Figure 5. Since a root function is isolated, any row can have at most four true minterms. Thus, if $N(R, 4) = 7$ for a root function, there must be four true minterms in one row and three in the other row. These four minterms must be placed in alternate cells in a row, which will dominate all false minterm in the row and four false minterm in the other row. Each of the remaining four cells in the row will be at Hamming distance 2 from each other, and hence three true minterms will be insufficient to dominate all the cells. Thus, no root function can exist with $N(R, 4) = 7$. □

| | 000 | 001 | 011 | 010 | 110 | 111 | 101 | 100 |
|---|---|---|---|---|---|---|---|---|
| 0 | | | | | | | | |
| 1 | | | | | | | | |

Figure 5: *K*-map in two rows for $n = 4$

Similarly, there cannot be any root function of 6 true minterms, as they can be placed either as 4 and 2 or as 3 and 3 in the two rows. By similar argument, one can show that no root function can exist with $N(R, 4) = 6$.

*Example 8*: An example of a root function for four variables with five true minterms is shown in Figure 6.

| $x_4x_3$ \ $x_2x_1$ | 00 | 01 | 11 | 10 |
|---|---|---|---|---|
| 00 | 1 | | | |
| 01 | | | 1 | |
| 11 | | 1 | | 1 |
| 10 | | | 1 | |

Figure 6: A root function for $n = 4$ with $N(R, 4) = 5$

## 2.2 Impossible class of faulty functions (ICFF)

In this subsection, we will show that the set of root functions plays an important role in partial characterization of ICFF in combinational circuits under stuck-at fault model (single/multiple). A combinational circuit is said to be *irredundant* of all faults are detectable at the output, else it is called *redundant*. A single-output two-level AND-OR circuit will be irredundant only if it implements an irredundant s-o-p expression of a Boolean function. Since, in a redundant circuit ICFF is unpredictable [15], we consider irredundant realizations only.

*Theorem 2*: Let $C$ be a single-output two-level irredundant combinational circuit realizing an *n*-variable Boolean function $F_0$. Let $F_f$ denote the faulty response function under any stuck-at fault in $C$. Then $F_f$ cannot be an *n*-variable root function.

*Proof:* In a two-level circuit, it is enough to consider stuck-at faults at the primary inputs (PI) and at the fanout branch lines only. If there is a fault at PI, $F_f$ becomes vacuous in some input literal and hence $F_f$ cannot be a root function. For any stuck-at 0 faults at inputs to AND gates, the corresponding prime implicant disappears from the output. For stuck-at 1 faults, they grow to product terms in which the corresponding literals are set to 1. Since $C$ is irredundant, in the first case, $F_f$ cannot be maximal and in the second case, $F_f$ cannot be isolated. Therefore, $F_f$ can never be a root function. □

*Theorem 3*: Let $F_f$ be a function of *n* variables. Then, there exists a root function $R(n)$ in whose irredundant two-level realization $C$, $F_f$ will appear as a faulty response on injection of some stuck-at faults in $C$.

*Proof*: Consider an irredundant s-o-p of $F_f$. It is always possible to choose a set $S$ of true minterms in $F_f$ such that their mutual Hamming distance is greater than 1, and which will grow to cover $F_f$ when certain literals are set to 0 or 1. Clearly, $S$ is an isolated function. If it is maximal, it is already a root function; if not, we add a set $M$ of few other minterms so that $S \cup M$ is maximal. Hence, a root function $R(n)$ exists. Thus, in $C$, certain stuck-at faults can be injected to produce $F_f$ as a faulty response. □

| $x_4x_3$ \ $x_2x_1$ | 00 | 01 | 11 | 10 |
|---|---|---|---|---|
| 00 | 1 | 1 | 1 | 1 |
| 01 | | | 1 | |
| 11 | | 1 | 1 | 1 |
| 10 | | | 1 | |

Figure 7: Illustration of Theorem 3

*Example:* Consider the function $F_f$ which is shown in Figure 7 as OR-ing of three prime implicants shown with green loops. We want to observe $F_f$ as a faulty response of a root function $R(n)$. To construct $R(n)$, we first pick up three isolated minterms shown in red circles from these three implicants, which can grow to realize $F_f$ under stuck-at 1 faults. To make it maximal, we add two more minterms shown in blue circles, such that no two minterms are adjacent. Thus, $R(n)$ will consist of sum of 5 minterms. This is a root function as illustrated earlier in Figure 6. In order to observe $F_f$ as a faulty response, consider an irredundant two-level implementation of $R(n)$. On injecting one stuck-at 0 fault (to cause disappearance of the minterm 1110) and few stuck at-1 faults (to cause growth of 0000 to 00--, 1011 to --11, and 1101 to 11-1), $F_f$ can be observed as a faulty response.

From Theorems 2 and 3, we have the following:
*Corollary*: The set of all root functions and the set of their possible faulty response functions under stuck-at faults are mutually exclusive and collectively exhaustive over all Boolean functions of *n* variables for any *n*, when single-output two-level irredundant circuit realizations are considered.

## 3. Relation to Independent Dominating Set

In this section, we show that the set of *n*-variable root functions has a one-to-one correspondence to the independent dominating sets in a hypercube of dimension *n* [17, 18].

### 3.1 Independent domination in hypercubes

Let $G = (V, E)$ denote a hypercube of dimension *n*, *i.e.*, $|V| = 2^n$ and there is an edge between two nodes in *G* if the Hamming distance between their binary representations is unity. A subset $S \in V$ is said to dominate *G* if every node in $V - S$ is adjacent to at least one node in *S*. If no two elements in *S* are adjacent in *G*, then *S* is called an independent dominating set. By $Q_{ind}(G)$ is meant the *minimum cardinality* of the independent dominating sets of *G*. Computation of minimum-size independent dominating set is a hard problem in graphs, although very good heuristic algorithms have recently been reported [21].

Since a Boolean function can be thought as a collection of 1-nodes (true minterms) in a hypercube, the following result is immediate.

*Theorem 4*: An *n*-variable Boolean function $R(n)$ is a root function if and only if $R(n)$ corresponds to an independent dominating set in a hypercube of dimension *n*. □

Thus, $Q_{ind}(G)$ represents the number of true minterms in a min-root function. It has been reported [17] that $Q_{ind}(G)$ is 1, 2, 2 4, 8, 12, for $n = 1, 2, 3, 4, 5, 6$ respectively, which match with our findings. However, till date not much is known about $Q_{ind}(G)$ for higher values of *n*. Also, enumeration of total number of root functions (or the number of independent dominating sets in a hypercube) is not yet studied. In the next subsection, we present our simulation results.

### 3.2 Experimental results

We have performed exhaustive enumeration of root functions for up to 6 variables and the results are reported in Table 1.

**Table 1:** Number of root functions for different numbers of true minterms in an *n*-variable function

| # of variables *n* | # true minterms in root functions | # of root functions | Total number of root functions |
|---|---|---|---|
| 2 | 2 | 2 | 2 |
| 3 | 2 | 4 | 6 |
| 3 | 4 | 2 | |
| 4 | 4 | 24 | 42 |
| 4 | 5 | 16 | |
| 4 | 8 | 2 | |
| 5 | 8 | 1140 | 1670 |
| 5 | 9 | 320 | |
| 5 | 10 | 176 | |
| 5 | 12 | 32 | |
| 5 | 16 | 2 | |
| 6 | 12 | 320 | 1,281,402 |
| 6 | 14 | 9600 | |
| 6 | 15 | 25920 | |
| 6 | 16 | 736440 | |
| 6 | 17 | 337920 | |
| 6 | 18 | 116320 | |
| 6 | 19 | 40320 | |
| 6 | 20 | 8320 | |
| 6 | 21 | 3840 | |
| 6 | 22 | 1856 | |
| 6 | 24 | 480 | |
| 6 | 27 | 64 | |
| 6 | 32 | 2 | |

## 4. Results and Discussion

Our experiments revealed many interesting results. For each value of *n*, we first check the possible number of true minterms in a root function. For certain integers *k* lying within the range of min-root size and $2^{n-1}$, a root function with *k* true minterms does not exist. For example, when $n = 5$, no root function exists for four values of *k*: 11, 13, 15, and 14. Similarly, for $n = 6$, no root functions exist with 13, 23, 25, 26, 28, 29, 30, 31 true minterms. For each *n*, we enumerate root functions with a possible number of true minterms (shown in the second column of the table). Also, all the root functions, which are *NP*-equivalent, are counted in one group of the table.

**Root functions as universal logic modules**

As a consequence of Theorem 3, the root functions can serve as universal logic modules for realizing any Boolean function. Given two-level realizations of root functions, all other non-root functions can be produced at the outputs on injection of certain stuck-at faults. We illustrate this principle for $n = 3$.

For three variables, there are six root functions (Table 1). These functions can be classifed into two NP-equivalence classes, which are shown in Figure 3(a) (consisting of two true minterms) and in Figure 3(b) (consisting of four minterms). We realize only these two functions by two-level AND-OR circuits, say $C_1$ and $C_2$ respectively. The remaining four root functions can be realized by permuting or complementing the input literals. We have checked that all the remaining 250 non-root functions can be produced from $C_2$ by injecting faults and by applying input permutations/literal complementation. However, 118 functions can be produced similarly from $C_1$ as well. Hence, the realizations of only two root functions are sufficient to produce all 256 functions of three variables. Similarly, three root functions are sufficient for producing all 65,536 four-variable functions.

## 4. Conclusion

We have introduced a new class of Boolean functions called root functions. The rationale behind naming them as "roots" lies in the fact that none of these functions can appear as faulty response in a single-output two-level irredundant circuit under any stuck-at fault. Also, they can produce any other non-root functions by suitable injection of faults in their irredundant realizations. Thus, these root functions can be used as universal logic modules to realize any Boolean function. We have shown that a root function is equivalent to an independent domination set in a hypercube and derived several new properties and bounds. We also computed the number possible root functions for up to six variables. There are many open problems related to this research, such as characterization of the inadmissible number of minterms in a root function and derivation of a closed formula for enumeration. Also, generalization of root functions and their fault behavior to more general circuits, which can be obtained from two-level circuits under certain transformations [19, 22] may be studied. Very recently, some authors are studying a special class of Boolean functions called "good functions", which are closely related to the notion of root functions [23].